\documentclass[a4paper,fleqn,usenatbib]{mnras}
\usepackage{newtxtext,newtxmath}
\usepackage[T1]{fontenc}
\usepackage{ae,aecompl}
\usepackage{graphicx}
\usepackage{amsmath}

\title[Elastic properties of multicomponent crystals in neutron stars and white dwarfs]{Elastic
properties of multicomponent crystals in neutron stars  and white dwarfs}
\author[A.~A. Kozhberov]{
A.~A. Kozhberov\thanks{E-mail:kozhberov@gmail.com} \\
Ioffe Institute, Politekhnicheskaya 26, 194021, Saint Petersburg, Russia}

\date{Accepted XXX. Received YYY; in original form ZZZ}

\pubyear{2019}

\begin{document}
\label{firstpage}
\pagerange{\pageref{firstpage}--\pageref{lastpage}}
\maketitle
\begin{abstract}
Elastic properties play an important role in neutron stars and white dwarfs. They are crucial for modeling stellar oscillations and different processes in magnetars and in degenerate stars which enter compact binary systems. Using electrostatic energy of deformed lattices, we calculate elastic moduli of ordered binary body-centered cubic (sc2) and face-centered cubic (fc2) lattices. We use two methods to determine the effective shear modulus $\mu_{\rm eff}$. We show that  $\mu_{\rm eff}$ calculated as a Voigt average agrees with the results obtained from the linear mixing rule. For the sc2 lattice, our calculations are also consistent with the results of numerical simulations of disordered binary body-centered cubic lattice.
\end{abstract}


\section{Introduction}
It is usually thought that outer crusts of neutron stars and cores of old white dwarfs consist of weakly polarized degenerate electrons and fully ionized atoms arranged into a crystal lattice \citep[e.g.][]{ST,HPY07}. This system could be described by the model of a Coulomb crystal of ions, where ions are treated as point particles while electrons form a uniform or weakly polarizable neutralizing background. Note that the inner crust of a neutron star contains also neutrons which weakly affect electron-ion interaction; the Coulomb crystal model can be extended for this part of the crust too. The model of the Coulomb crystal allows one to study thermodynamic \citep[e.g.][]{B01}, transport  \citep[e.g.][]{PB99} and others properties of stellar matter \citep[e.g.][]{CH08}. In this paper, we focus on static elastic properties of Coulomb  solids in the interiors of degenerate stars.

The study of elastic properties of the crust is a very important issue in physics of neutron stars especially in physics of their oscillations. \citet{S91} demonstrated that toroidal, spheroidal and interfacial oscillation modes significantly depend on the effective shear modulus $\mu_{\rm eff}$. In turn, toroidal modes are used for the interpretation of global seismic oscillations of soft gamma repeaters \citep[e.g.][]{D98} and quasi-periodic oscillations of magnetars \citep[e.g.][]{SW06,G18}. According to some models, the magnetar activity is generated by shear motions near the neutron star surface and therefore depends on $\mu_{\rm eff}$ \citep{BL14,LBL16}. The problem of mountain formation on the neutron stars surface is directly related to the elastic properties. These mountains can be efficient sources of gravitational waves \citep[e.g.][]{U00,H06,JO13,HP17,A19}. Hence, the effective shear modulus is required to interpret different sets of observational data. In this way it serves as an important microphysical parameter of neutron star envelopes.

Many previous publications considered lattices in neutron stars and white dwarfs as one-component. In addition it was often  thought that the ions form the body-centered cubic (bcc) Coulomb lattice \citep[e.g.][]{CF16}. The elastic moduli of the bcc lattice were investigated in several works \citep{F36,W67,RKG88,OI90,S91,II03,HH,B11,B15} both numerically and analytically. For instance,  \citet{OI90,S91} employed the Monte Carlo method and analyzed free energy changes during lattice deformations. In this approach, elastic moduli of the static lattice are determined through a limit of free energy changes at $T=0$.

However, investigations of multi-component crystals are certainly important, particularly, for compact binaries (containing a neutron star or/and a white dwarf). These systems are interesting not only as the sources of gravitational waves but also as the objects for studying intensive accretion processes. Simulations show that neutron star crust in a compact binary can be very heterogeneous as far as its ion composition is concerned \citep[e.g.][]{DG09,HB09,CCB18}.

Note that the cores of white dwarfs are composed of carbon -- oxygen mixture with traces of other elements \citep[e.g.][]{S94}. During the thermal evolution, such a mixture crystallizes with the formation of a multi-component Coulomb crystal \citep[e.g.][]{A10}.

Previously, elastic properties of multi-component Coulomb compounds have been studied, as far as we know, only numerically for binary lattices in the disordered state (i.e. neglecting correlations in positions of different ions) by \citet{II03}.

In the present paper, we calculate the elastic properties of one-component and binary ordered Coulomb lattices. We use the zero temperature limit and consider only static elastic properties. It is a good approximation for internal neutron star temperatures below $10^8$ K \citep[e.g.][]{G11}. We examine two different approaches to calculate  $\mu_{\rm eff}$ \citep[see][for a  review]{KP15} and two types of crystal lattices. We discuss also the applicability of the linear mixing rule to determine the elastic moduli. The influence of the finite temperature to the multi-component systems may be valuable for young neutron stars and could be studied via thermodynamic perturbation theory \citep{B11,B15} or molecular-dynamic simulations \citep{HH}.

The present paper is organized as follows. Section \ref{ele} discusses electrostatic energies of a  binary bcc lattice stretched along the edges of the basic lattice cube. These energies are used to calculate the elastic coefficients of the binary bcc lattice in Section \ref{elc} and the effective shear modulus in Section \ref{esm}. The importance of screening corrections is discussed in Section \ref{sc}. The elastic coefficients and the effective shear modulus of the binary face-centered cubic lattice are studied in Section \ref{bfc}. Astrophysical implications are outlined in Section \ref{dc}.


\section{Electrostatic energy}
\label{ele}
Consider the simplest case of multi-component Coulomb crystals --- an ordered binary body-centered cubic (sc2) lattice. Following  \citet{KB12,KB151}, we describe it as a simple cubic lattice with two ions in the elementary cell ($N_{\textrm{cell}}=2$). So we use the main translation vectors for the sc2 lattice, $\textbf{a}_1=a_\textrm{l}(1,0,0)$, $\textbf{a}_2=a_\textrm{l}(0,1,0)$,  $\textbf{a}_3=a_\textrm{l}(0,0,1)$, the basis vectors $\boldsymbol{\chi}_1=0$, $\boldsymbol{\chi}_2 = 0.5 a_\textrm{l} (1,1,1)$, where $a_\textrm{l}$ is the lattice constant. The basis vector $\boldsymbol{\chi}_1$ corresponds to the ion with the charge number $Z_1$ and the basis vector $\boldsymbol{\chi}_2$ corresponds to the ion with the charge number $Z_2$. The origin of a Cartesian coordinate system is chosen such that $Z_2 \geq Z_1$. The number density of ions with the charge number $Z_1$ is denoted as $n_1$, and the number density of other ions is $n_2$ (in the sc2 lattice $n_1=n_2$).

According to  \citet{KB12,KB151}, the electrostatic energy of the ordered multi-component lattice can be written as
\begin{eqnarray}
U_\textrm{M} &=& N\frac{Z_1^2e^{2}}{a}\xi~,
\nonumber \\
\xi&=&\frac{a}{2N_\textrm{cell}}\sum_{lpp'}\frac{{Z}_{p}{Z}_{p'}}{Z_1^2}
  \left(1-\delta_{pp'}\delta_{\textbf{R}_l0}\right) \frac{\textrm{erfc}
  \left(A Y_{lpp'}\right)}{Y_{lpp'}}
\nonumber \\
&-&\frac{Aa}{N_\textrm{cell}\sqrt{\pi}}\sum_p \frac{{Z}_{p}{Z}_{p'}}{Z_1^2}
  -\frac{3}{8N_\textrm{cell}^2A^2a^2}\sum_{pp'}\frac{{Z}_{p}{Z}_{p'}}{Z_1^2}
\nonumber \\
&+&\frac{3}{2N_\textrm{cell}^2 a^2}\sum_{mpp'}\frac{{Z}_{p}{Z}_{p'}}{Z_1^2}
    (1-\delta_{\textbf{G}_m0})
\nonumber \\
&\times& \frac{1}{G_m^2}\exp\left[-\frac{G_m^2}{4A^2}+
     i\textbf{G}_m(\boldsymbol{\chi}_p
      -\boldsymbol{\chi}_{p'})\right]~,
\label{Mad}
\end{eqnarray}
where ${\textbf{Y}}_{lpp'}={\textbf{R}}_l+\boldsymbol{\chi}_p-\boldsymbol{\chi}_{p'}$, ${\bf R}_l=l_1 \textbf{a}_1+l_2 \textbf{a}_2+l_3 \textbf{a}_3$ are the lattice vectors, $l_1, l_2, l_3$ are arbitrary integers, $\textbf{G}_m=m_1 \textbf{g}_1+m_2 \textbf{g}_2+m_3 \textbf{g}_3$ are the vectors of reciprocal lattice, $\textbf{g}_{i}\textbf{a}_{j}=2\pi\delta_{ij}$, $m_1, m_2, m_3$ are arbitrary integers, sums over $p$ and $p'$ go over all ions in the elementary cell (in the ordered crystal the charge number of an ion $Z_p$ depends only on its place in the elementary cell), ${\rm erfc}(x)$ is the complementary error function, $N$ is the total number of ions, $a\equiv(4\pi n/3)^{-1/3}$ is the ion sphere radius, and $n$ is the total number density of ions (for the sc2 lattice, $n=n_1+n_2=2n_1$). $A$ is an arbitrary parameter ($Aa\approx1$), which is chosen to optimize the convergence of summation in Eq. (\ref{Mad}) and $U_{\rm M}$ is independent of $A$.

The parameter $\xi$ is the Madelung constant. For any binary crystal it depends only on $\alpha\equiv Z_2/Z_1$. For the sc2 lattice
\begin{eqnarray}
\xi_{\rm sc2}&=&\frac{1+\alpha^{2}}{2^{4/3}}
\xi_1+\alpha\left(\xi_2-\frac{\xi_1}{2^{1/3}}\right) \nonumber \\
&=&-0.3492518\left(1+\alpha^{2}\right)-0.1974256\alpha~,
\label{E2}
\end{eqnarray}
where $\xi_1=\xi_{\rm sc}=-0.88005944211$ is the Madelung constant of the simple cubic (sc) lattice and $\xi_2=\xi_{\rm bcc}=-0.89592925568$ is the Madelung constant of the bcc lattice.

Eq. (\ref{Mad}) allows one to calculate the electrostatic energy of any periodic multi-component lattice, in particular the sc2 lattice with arbitrary uniform deformation. In this section, we consider the sc2 lattice stretched along edges of the main lattice cube in such a way  that the main translation vectors tend to $\textbf{a}_1=a_\textrm{l}(1,0,0)$, $\textbf{a}_2=a_\textrm{l}(0,c_1,0)$, $\textbf{a}_3=a_\textrm{l}(0,0,c_2)$, where the parameters $c_1$ and $c_2$ characterize the stretch value.
In the common case the volume of the elementary cell changes during the stretching ($na^3_{\rm l}=2/(c_1c_2)$ at arbitrary $c_1$ and $c_2$). It is stay constant only at $c_2=1/c_1$.

The Madelung constant $\xi(c_1,c_2)$ of the stretched sc2 lattice  depends on $\alpha$, $c_1$ and $c_2$. For $1.0 \leq c_1, c_2 \leq 1.6$ it can be approximated as
\begin{eqnarray}
\xi(c_1,c_2)&=&K_1(c_1, c_2)(1+\alpha^2)+K_2(c_1, c_2)\alpha, \\
K_1(c_1, c_2)&=&\sum_{i=0}^4 m_i (c_1^i+c_2^i)+n_1c_1c_2\nonumber \\
&+&n_2c_1^2 c_2^2+n_3(c_1^2c_2+c_1c_2^2)+n_4(c_1^3 c_2+c_1c_2^3). \nonumber \\
K_2(c_1, c_2)&=&\sum_{i=0}^4 x_i (c_1^i+c_2^i)+y_1c_1c_2\nonumber \\
&+&y_2c_1^2 c_2^2+y_3(c_1^2c_2+c_1c_2^2)+y_4(c_1^3 c_2+c_1c_2^3). \nonumber
\end{eqnarray}
The parameters of the approximation $m_i$ and $n_i$ are presented in Table \ref{tab:apr2}. Errors of the approximation do not exceed 0.02\%.

\begin{table}
	\centering
	\caption{Parameters of the approximation of the Madelung constant of the stretched sc2 lattice.}
	\label{tab:apr2}
	\begin{tabular}{ccccc}
	\hline
& \multicolumn{2}{c}{$K_1(c_1,c_2)$} & \multicolumn{2}{c}{$K_2(c_1,c_2)$} \\
\hline
$i$ & $m_i$ & $n_i$ & $x_i$ & $y_i$\\
\hline
0 & $-0.0616385$ &   & $-0.3722545$ & \\
1 & $-0.301618$ & $-0.3857925$  & 0.776568 & 0.010696 \\
2 & 0.474446 & 0.0428725 & $-0.793227$ & 0.021387 \\
3 & $-0.202857$ & 0.0924215 & 0.240636 & 0.10033 \\
4 & 0.041644 & $-0.045568$ & $-0.021922$ & $-0.04487$ \\
\hline
	\end{tabular}
\end{table}

At any fixed $Z_1$ and $Z_2$ the electrostatic energy of the stretched sc2 lattice reaches its minimum at $c_1=1$ and $c_2=1$, so that the bcc lattice does not change its shape when $\alpha$ changes. Note that in the binary hexagonal close packed lattice the distance between hexagonal layers decreases with increasing $\alpha$ \citep{K18}.


\section{Elastic coefficients}
\label{elc}
In some cases it is convenient to rewrite the Madelung constant as
\begin{equation}
\xi'\equiv\xi \frac{2a_{\rm l}}{a}~.
\end{equation}
For the sc2 lattice, $na^3_{\rm l}=2$; hence, $a^3_{\rm l}=8\pi a^3/3$ and
\begin{equation}
\xi'_{\rm sc2}=4(\pi/3)^{1/3}\xi_{\rm sc2}~.
\end{equation}
For the stretched sc2 lattice,
\begin{equation}
\xi'(c_1,c_2)= 4\left(\frac{\pi}{3c_1c_2}\right)^{1/3}\xi(c_1,c_2)~.
\end{equation}

If lattice deformations are small, we can expand $\xi'(c_1,c_2)$ in powers of $c_1-1$ and $c_2-1$,
\begin{eqnarray}
\xi'(c_1,c_2) &\approx& \xi'_{\rm sc2}-\tilde{p}'_{\rm sc2}\left[(c_1-1)+(c_2-1)\right]\nonumber \\
&+&0.25 \tilde{s}^{xxxx}_{\rm sc2}\left[(c_1-1)^2+(c_2-1)^2\right] \nonumber \\
&+&0.5\tilde{s}^{xxyy}_{\rm sc2}(c_1-1)(c_2-1)~,
\label{SM2}
\end{eqnarray}
where $\tilde{p}'_{\rm sc2}=\xi'_{\rm sc2}/3$ is the electrostatic pressure (this equality is valid for any lattice with the isotropic pressure). The parameters $\tilde{s}^{xxxx}_{\rm sc2}$ and $\tilde{s}^{xxyy}_{\rm sc2}$ are the static lattice elastic coefficients
\begin{eqnarray}
\tilde{s}^{xxxx}_{\rm sc2}&=&0.32969383(1+\alpha^2)-2.144195558\alpha~, \label{sxx} \\
\tilde{s}^{xxyy}_{\rm sc2}&=&-0.637729828(1+\alpha^2)+0.804785789\alpha~. \label{sxy}
\end{eqnarray}

At $\alpha=1$ the sc2 lattice tends to the bcc lattice. Hence $\tilde{s}^{xxxx}_{\rm bcc}=-1.48480792$ and $\tilde{s}^{xxyy}_{\rm bcc}=-0.47067387$. For the bcc lattice these elastic coefficients were obtained earlier \citep{F36,OI90,B11}. \citet{B11}  denoted them as $S^{\rm st}_{1111}$ and $S^{\rm st}_{1122}$, respectively, and our results fully reproduce them. Notice that $\tilde{s}^{xxxx}+2\tilde{s}^{xxyy}=2\tilde{p}'$ for any cubic Coulomb lattice \citep{Ch}.

The ideal cubic crystal lattice has three independent elastic moduli. Two of them are $\tilde{s}^{xxxx}$ and $\tilde{s}^{xxyy}$. The third static lattice elastic modulus can be found from the analysis of the electrostatic energy of the sc2 lattice with a shift. In the elementary cell of this lattice the top
layer is horizontally shifted with respect to the bottom layer (cube of the elementary cell turns to a square based prism). The main translation vectors of the sc2 lattice with a shift can be defined as $\textbf{a}_1=a_\textrm{l}(1,0,0)$, $\textbf{a}_2=a_\textrm{l}(0,1,0)$, $\textbf{a}_3=a_\textrm{l}(c_x,c_y,1)$. In this case the volume of the elementary cell does not change during deformation and $na^3_{\rm l}=2$.

The Madelung constant of the sc2 lattice with a shift $\xi'(c_x,c_y)$ depends on $\alpha$ and on parameters of deformation $c_x$ and $c_y$. At small $c_x$ and $c_y$ the quantity $\xi'(c_x,c_y)$  can be written as
\begin{eqnarray}
\xi'(c_x,c_y)&\approx&\xi'_{\rm sc2}+0.25\tilde{s}^{xyxy}_{\rm sc2}\left(c_x^2+c_y^2\right)~, \label{nak2} \\
\tilde{s}^{xyxy}_{\rm sc2}&=&-0.164846915(1+\alpha^2)+1.072097779\alpha. \label{sxyx}
\end{eqnarray}
For the bcc lattice, we have $\tilde{s}^{xyxy}_{\rm bcc}=0.74240395$, which agrees with the result of \cite{B11}  where this coefficient is denoted as $S^{\rm st}_{1212}$. One can see that, numerically, $\tilde{s}^{xyxy}_{\rm sc2}=-\tilde{s}^{xxxx}_{\rm sc2}/2$.

In Eqs. (\ref{sxx},\ref{sxy},\ref{sxyx}) the static lattice elastic coefficients and the pressure are presented in dimensionless units. In physical units we can write that for the sc2 lattice
\begin{eqnarray}
s^{xyxy}_{\rm sc2}&=&n\frac{Z_1^2e^2}{2a_{\rm l}} \tilde{s}^{xyxy} \\
&=&n\frac{Z_1^2e^2}{2a_{\rm l}}\left(-0.164846915(1+\alpha^2)+1.072097779\alpha\right)~. \nonumber
\end{eqnarray}
In some papers \citep[e.g.][]{OI90,B15} the elastic coefficients are measured in units of $a$ instead of $2a_{\rm l}$. Then, for instance,
\begin{equation}
c_{44}\equiv s^{xyxy} \frac{a}{2a_{\rm l}}~, \quad c_{44, \rm bcc}=0.18276965 n\frac{Z_1^2e^2}{a}~.
\end{equation}

Previously, elastic coefficients of multi-component lattices have been studied only via molecular-dynamic simulations by \citet{II03}; and only the disordered binary bcc lattice has been considered (in the disordered lattice, ions of different charge are randomly distributed along lattice cites). In the sc2 lattice, we have $n_1=n_2$, while in a disordered binary bcc lattice the $n_1/n_2$ ratio can be arbitrary.
In \citet{II03} elastic coefficients were calculated by taking the second derivative of the Madelung energy, while the Madelung energy was defined as a minimal energy of around 1000 ions in the cubic cell (the minimization procedure was iterated until the relative variance of the energy reached $10^{-6}$).

It is well known that the linear mixing (lm) rule has been successfully applied for calculating thermodynamic properties of classical Coulomb mixtures \citep[e.g.][]{ChA90,P09,C12} and the electrostatic energy of the sc2 lattice \citet{KB151}. Here we try to check its validity for the elastic moduli of Coulomb lattices. According to this rule, the modulus $c_{44}$ of any (ordered and disordered) binary lattice is equal to
\begin{equation}
c_{44}^{\rm lm}\equiv c_{44}^{Z_1}\left(\frac{n_1}{n}+\frac{n_2}{n}\alpha^{5/3}\right)\left(\frac{n_1}{n}+\frac{n_2}{n}\alpha\right)^{1/3}~,
\end{equation}
where $c_{44}^{Z_1}$ is the elastic modulus of the one-component lattice which consist of ions with the charge number $Z_1$. For the sc2 lattice,
\begin{equation}
c_{44,\rm sc2}^{\rm lm}=0.18276965n\frac{Z_1^2e^2}{2^{4/3}a}(1+\alpha^{5/3})(1+\alpha)^{1/3}~.
\end{equation}
Other elastic moduli can be calculated in the same way.

Thus the elastic coefficients of the binary bcc lattice with $n_1=n_2$ can be calculated analytically from the electrostatic energy of the deformed crystal via linear mixing rule and obtained from numerical simulations. For several values of $\alpha$ the coefficients $c_{44}$ are presented in Table \ref{tab:mod1}. The results of \cite{II03} are labeled as $c_{44}^{\rm dis}$. \cite{II03} investigated lattices with $\alpha \leq 13$ while we restrict ourselves to $\alpha=3$ because (as shown by \citealt{KB12}) the sc2 lattice is stable with respect to the small oscillations of ions around their equilibrium positions at $\alpha<3.6$ (similar result was obtained latter in \citet{KD14} from molecular dynamics).

\begin{table}
	\centering
	\caption{Values of $c_{44}$ in units of $nZ_1^2e^2/a$ for the binary bcc lattice.}
	\label{tab:mod1}
	\begin{tabular}{cccc}
\hline
$\alpha$ & $4/3$ & 2 & 3 \\
\hline
$c_{44,\rm sc2}$ & 0.239184 & 0.324956 & 0.385977 \\
$c_{44,\rm sc2}^{\rm dis}$  & 0.241  & 0.292 & 0.549 \\
$c_{44,\rm sc2}^{\rm lm}$  & 0.25159 & 0.43672 & 0.83363 \\
\hline
\end{tabular}
\end{table}

One can see a noticeable difference between $c_{44}$ for ordered and disordered binary bcc lattices, especially at high $\alpha$. The linear mixing rule does not allow one to calculate this elastic coefficient with appropriate precision. The same is true for other elastic moduli ($s^{xxxx}$ and $s^{xxyy}$) of the sc2 lattice.


\section{Effective shear modulus}
\label{esm}
\begin{table}
	\centering
	\caption{Values of ${\mu}_{\rm eff}$ in units of $nZ_1^2e^2/a$ of the binary bcc lattice.}
	\label{tab:mod2}
	\begin{tabular}{cccc}
\hline
$\alpha$ & $4/3$ & 2 & 3 \\
\hline
$\mu_{\rm eff}^{\rm sc2}$ & 0.164451 & 0.285482 & 0.544639 \\
$\mu_{\rm eff}^{\rm dis}$ & 0.164 & 0.284 & 0.542  \\
$\mu_{\rm eff}^{\rm lm,sc2}$ & 0.164439 & 0.28544 & 0.544853 \\
$\mu_{\rm eff,m}^{\rm sc2}$ & 0.139547 & 0.282088 & 0.517677 \\
$\mu_{\rm eff,m}^{\rm dis}$ & 0.138 & 0.284 & 0.542  \\
$\mu_{\rm eff,m}^{\rm lm,sc2}$ & 0.128027 & 0.222235 & 0.424206 \\
\hline
\end{tabular}
\end{table}
The matter in a neutron star crust is often assumed to be polycrystalline \citep[e.g.][]{HK09,CCB18}. Then the crust consist of randomly oriented crystals, and it is convenient to use an effective shear modulus $\mu_{\rm eff}$. However, the question of how to determine $\mu_{\rm eff}$ remains open \citep{KP15}.

At first, we use the way, that is the most common in theory of degenerated stars, and define the effective shear modulus as an average of the shear stiffness over all possible wavevectors and for polarisation vectors perpendicular to the wave vector, which is equivalent to the Voigt average (see details in \citealt{OI90,B11} for the one-component crystals). In our notations,
\begin{equation}
\mu_{\rm eff}\equiv \frac{1}{5}(s^{xxxx}-s^{xxyy}+3s^{xyxy}-p')~. \label{mu}
\end{equation}

For the sc2 lattice, we have $s^{xyxy}_{\rm sc2}=-s^{xxxx}_{\rm sc2}/2$, $s^{xxxx}_{\rm sc2}+2s^{xxyy}_{\rm sc2}=2p'$, and $\tilde{p}'=\xi'/3$. Then $\mu_{\rm eff}^{\rm sc2}$ depends only on the Madelung constant  \citep{Ch},
\begin{eqnarray}
\mu_{\rm eff}^{\rm sc2}&=&-\frac{2}{15}\xi_{\rm sc2}\frac{nZ_1^2e^2}{a} \nonumber \\
&=&\left(0.0465669(1+\alpha^2)+0.0263234\alpha\right)\frac{nZ_1^2e^2}{a}~.
\label{msc2}
\end{eqnarray}
For the bcc lattice
\begin{eqnarray}
\mu_{\rm eff}^{\rm bcc}&=&-\frac{2}{15}\xi_{\rm bcc}n\frac{Z^2e^2}{a} \nonumber \\
&=&0.1194572n\frac{Z^2e^2}{a}=0.4852310n\frac{Z^2e^2}{2a_{\rm l}}~,
\end{eqnarray}
which agrees with the results of \citet{B11}.

According to \cite{KB151}, the linear mixing rule can be successfully used to calculate the electrostatic energy of the sc2 lattice ($U_{\rm M}^{\rm lm}$). The ratio of the exact  $U_{\rm M}$ to $U_{\rm M}^{\rm lm}$ lies between 0.999843 at $\alpha \approx 1.82619$ and 1.00094 at $\alpha=3.6$. Hence the linear mixing rule can also be successfully applied to determine the effective shear modulus defined by Eq. (\ref{mu}),
\begin{equation}
{\mu}_{\rm eff}^{\rm lm,sc2}=0.119457n\frac{Z_1^2e^2}{2^{4/3}a}(1+\alpha^{5/3})(1+\alpha)^{1/3}~.
\end{equation}
For some values of $\alpha$, the quantities  $\mu_{\rm eff}^{\rm sc2}$ and $\mu_{\rm eff}^{\rm lm,sc2}$ are presented in Table  \ref{tab:mod2}.

The effective shear modulus of the disordered binary bcc lattice obtained by \citet{II03} is denoted in Table \ref{tab:mod2} as $\mu_{\rm eff}^{\rm dis}$. The difference between $\mu_{\rm eff}$ for the ordered and disordered lattices is insignificant and does not exceed computational errors.

\begin{figure}
	\includegraphics[width=\columnwidth]{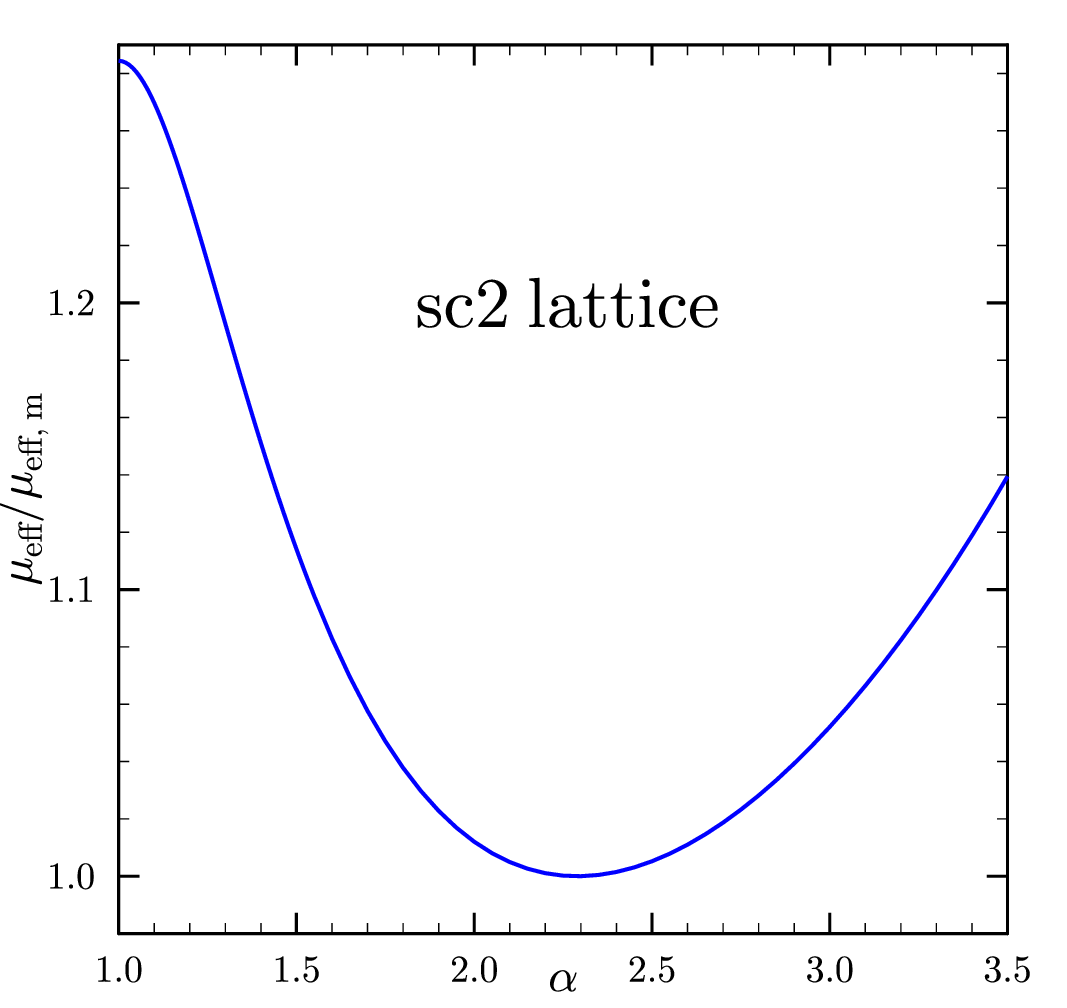}
    \caption{The ratio of the
    	 $\mu_{\rm eff}$ values calculated by different methods for the sc2 lattice.}
    \label{fig:mod2}
\end{figure}

There are some other ways to determine the effective shear modulus of polycrystals. For the bcc lattice, they are summarized  by \cite{KP15}. Among all of these ways we focus on the effective medium theory based on a multiple scattering formalism \citep[e.g.][]{Z73,RKG88}. According to this theory, the effective elastic modulus of a one-component cubic crystal can found from the equation,
\begin{equation}
3{\mu}_{\rm eff,m}^2 -s^{xyxy}{\mu}_{\rm eff,m}-(s^{xxxx}-s^{xxyy}-p')s^{xyxy}=0~,
\label{mum}
\end{equation}
where we take into account that the dominated contribution to the pressure comes from degenerated elections \citep{KP15}. For the bcc lattice we obtain
\begin{equation}
\mu_{\rm eff,m}^{\rm bcc}=0.377786n\frac{Z^2e^2}{2a_{\rm l}}=0.0930057n\frac{Z^2e^2}{a}~.
\end{equation}
For the sc2 lattice and several values of $\alpha$, the values of $\mu_{\rm eff,m}^{\rm sc2}$ are given in Table \ref{tab:mod2}. For this definition of the effective shear modulus, exact results are inconsistent with the results of numerical simulations (\citealt{II03};  $\mu_{\rm eff,m}^{\rm dis}$ in Table \ref{tab:mod2}) and with the results obtained from the linear mixing rule ($\mu_{\rm eff,m}^{\rm lm,sc2}$ in Table  \ref{tab:mod2}).
The difference between $\mu_{\rm eff}^{\rm sc2}$ and $\mu_{\rm eff,m}^{\rm sc2}$ is appreciable. The  $\mu_{\rm eff}^{\rm sc2}/\mu_{\rm eff,m}^{\rm sc2}$ ratio changes nonmonotonically with $\alpha$ as plotted in Fig. \ref{fig:mod2}. At $\alpha=1$ it is equal 1.28441 which is the maximum. The minimum is reached at $\alpha \approx 2.29$ and equals 1. At $\alpha=3.6$ we have $\mu_{\rm eff}^{\rm sc2}/\mu_{\rm eff,m}^{\rm sc2} \approx 1.16175$.
Notice that at $\alpha=2$ and $\alpha=3$ for the disordered crystal both methods give the same result  ($\mu_{\rm eff}^{\rm dis}=\mu_{\rm eff,m}^{\rm dis}$).


\section{Screening corrections}
\label{sc}
Using the same method as in the previous sections, we can calculate the elastic coefficients of the one-component bcc Coulomb crystal with polarized electron background. The polarization correction to the electrostatic energy was derived by \citet{B02} (see his Eq. (9)). The only difference from the uniform case is that the screening parameter $\kappa_{\rm TF}a$ also depends on volume changes during stretches along edges of the main lattice cube; for degenerate electrons and one-component lattices $\kappa_{\rm TF}a \approx0.185Z^{1/3}(1+x_{\rm r}^2)^{1/4}/x_{\rm r}^{1/2}$, where $x_{\rm r}$ is the electron relativity parameter.

Screening corrections (scr) to the electrostatic pressure and effective shear modulus of the bcc lattice are
\begin{eqnarray}
p^{\rm scr}&=&\frac{\eta}{3} \frac{x_{\rm r}^2}{1+x_{\rm r}^2}(\kappa_{\rm TF}a)^2n\frac{Z^2e^2}{a} \\
c^{\rm scr}_{44}&=&-0.041198(\kappa_{\rm TF}a)^2n\frac{Z^2e^2}{a} \\
\mu_{\rm eff}^{\rm scr}&\approx&-0.027662(\kappa_{\rm TF}a)^2n\frac{Z^2e^2}{a} \nonumber \\
&=&\frac{4\eta}{15}(\kappa_{\rm TF}a)^2n\frac{Z^2e^2}{a}~,
\end{eqnarray}
where $\eta=-0.1037323337$ is the screening correction to the electrostatic energy. These results agree with and improve the results of \citet{B15}, where the screening corrections to the Coulomb crystal elastic moduli were systematically studied for the first time. Also, they were studied by \citet{HH} via molecular dynamics simulations for the bcc Coulomb crystal. At $\kappa_{\rm TF}a \approx 0.5705$ and $T=0$ it was obtained that the effective shear modulus defined by Eq. (\ref{mu}) equals $0.1108nZ^2e^2/a$. Our calculations give $\mu_{\rm eff} \approx 0.1105nZ^2e^2/a$, so that the agreement is quite well. Note that \citet{RKG88} studied the screening corrections for the systems with nondegenerate electron background. For that reason, the direct comparison with our results is not possible.


\section{Binary face-centered cubic lattice}
\label{bfc}
We have also considered the binary fc2 lattice. In this case $n_2=3n_1$. The ions with lower
number density have the change number $Z_1$ (see Fig. 3 from \citealt{CF16} where
this lattice is called `the sc2 lattice'). In the elementary cell ions with $Z_1$ located on corners of the cube, ions with $Z_2$ are centered on its faces. The electrostatic energy of the fc2 lattice is
\begin{eqnarray}
U_{\rm M}&=&-N\frac{Z_1^2 e^2}{2a_{\rm l}}\xi'_{\rm fc2}=-N\frac{Z_1^2 e^2}{a}(0.138600677 \nonumber\\
&+&0.1707354535\alpha+0.5865374846\alpha^2)~,
\label{bin_fcc}
\end{eqnarray}
and $\xi'_{\rm fc2}=2(2\pi/3)^{1/3}\xi_{\rm fc2}$. For the first time this expression was obtained by \citet{J82}; here it is presented with improved  accuracy.

The consideration of the same deformations as for the sc2 lattice in Section \ref{elc} gives
\begin{eqnarray}
\tilde{p}'_{\rm fc2}&=&\xi'_{\rm fc2}/3 \label{pf} \\
\tilde{s}^{xxxx}_{\rm fc2}&=&0.16484692 \nonumber \\
&-&1.27801856\alpha-0.78347781\alpha^2~, \label{s1f} \\
\tilde{s}^{xxyy}_{\rm fc2}&=&-0.31886491 \nonumber \\
&+&0.34774851\alpha-0.60884623\alpha^2~,  \label{s2f} \\
\tilde{s}^{xyxy}_{\rm fc2}&=&-\tilde{s}^{xxxx}_{\rm fc2}/2 \label{s3f}~.
\end{eqnarray}

For the fcc lattice we obtain $\tilde{p}'_{\rm fcc}=-1.528287358$, $\tilde{s}^{xxxx}_{\rm fcc}=-1.89664945$, $\tilde{s}^{xxyy}_{\rm fcc}=-0.57996263$ and $\tilde{s}^{xyxy}_{\rm fcc}=0.94832473$, then $-\tilde{p}'_{\rm fcc}+\tilde{s}^{xxxx}_{\rm fcc}-\tilde{s}^{xxyy}_{\rm fcc}=0.21160053$. The latter value is in good agreement with the result of \cite{F36} who obtained 0.2115. The values $\tilde{s}^{xxxx}_{\rm fcc}$ and $\tilde{s}^{xxyy}_{\rm fcc}$ were not presented by \cite{F36}. According to  \cite{F36}, $\tilde{s}^{xyxy}_{\rm fcc}=0.9479$ which also agrees with our result. The one-component fcc lattice was also studied by \cite{RKG88}.

From Eqs. (\ref{pf}--\ref{s3f}) the effective shear modulus of the fc2 lattice defined by Eq. (\ref{mu}) is
\begin{eqnarray}
\mu_{\rm eff}^{\rm fc2}&=&-\frac{2}{15}\xi_{\rm fc2}n\frac{Z_1^2e^2}{a}=n\frac{Z_1^2e^2}{a}\left(0.01848009 \right. \nonumber \\
&+&\left.0.02276473\alpha+0.07820500\alpha^2\right)~.
\label{mfc2}
\end{eqnarray}

The effective shear modulus of the fcc lattice ($\alpha=1$) is equal to
\begin{equation}
\mu_{\rm eff}^{\rm fcc}=-\frac{2}{15}\xi_{\rm fcc}n\frac{Z^2e^2}{a}=0.11944982n\frac{Z^2e^2}{a}~.
\end{equation}
Data from \cite{F36} gives the same value. The difference between $\mu_{\rm eff}^{\rm bcc}$ and $\mu_{\rm eff}^{\rm fcc}$ is small but only for the chosen definition of $\mu_{\rm eff}$ because the difference between the Madelung constants of these lattices is small. In this case
\begin{equation}
\mu_{\rm eff,m}^{\rm fcc}=0.0901087n\frac{Z^2e^2}{a}~.
\end{equation}
Note that the bcc lattice can be turned into the fcc lattice by continuous deformation \citep{BK17}.

According to the linear mixing rule, ${\mu}_{\rm eff}$ of the binary fcc lattice is
\begin{equation}
{\mu}_{\rm eff}^{\rm lm,fc2}=0.11944982n\frac{Z_1^2e^2}{2^{8/3}a}(1+3\alpha^{5/3})(1+3\alpha)^{1/3}~.
\end{equation}
Our analysis of the phonon spectrum shows that the fc2 lattice is stable at $0.66 \leq \alpha \leq 1.36$
\citep[step over $\alpha$  equals 0.02; see ][for details]{K18}. This result agrees with the limits of stability of the fc2 lattice obtained
independently by \citet{KD14}: $0.661 \leq \alpha \leq 1.368$. For this range of $\alpha$, the ${\mu}_{\rm eff}^{\rm lm,fc2}/\mu_{\rm eff}^{\rm fc2}$ ratio always ranges between 1 and 1.002.
As for the sc2 lattice, the elastic moduli $s^{xxxx}$, $s^{xxyy}$ and $s^{xyxy}$ of the fc2 lattice cannot be calculated using the linear mixing rule.


\section{Discussion and conclusions}
\label{dc}
We discuss elastic properties of the binary Coulomb crystals. Our results demonstrate that the Voigt averaged effective shear modulus calculated for ordered crystals well agrees with the numerical results for disordered crystals. It is also shown, that the linear mixing rule can be applied to calculate $\mu_{\rm eff}$, providing thus a simple approach to estimate effective shear modulus for neutron star crust and crystallized white dwarf core. It should be stressed that the linear mixing rule is inapplicable to other elastic moduli of the sc2 and fc2 latices. The possible explanation may be that $s^{xxxx}$, $s^{xxyy}$ and $s^{xyxy}$ related with properties of the separated parts of the crystal and not with the whole one. The same concern the difference between our results and results obtained in \citet{II03}. This discrepancy can be a great motivation for the future numerical work. Furthermore, it is shown that the elastic constants for sc2 and fc2 lattices have additional coupling, which do not follows from their symmetry: $s^{xxxx}=-2\,s^{xyxy}$.

The Voigt average is not the only approach to estimate the effective shear modulus of the polycrystalline matter (see Sec. \ref{esm}), in particular it can be estimated according to Eq. (\ref{mum}). Here we show that the resulting $\mu_{\rm eff,m}$ differs for ordered and disordered crystals.
For ordered crystals ratio between $\mu_{\rm eff}$ and $\mu_{\rm eff,m}$ is less than 30\% for any possible charge ratio. In addition, the linear mixing rule does not allow to calculate $\mu_{\rm eff,m}$ with the appropriate for the practical use precision. Applying obtained results we can conclude that for the binary systems in degenerated stars it may be better to use the Voigt averaged effective shear modulus because it contains less uncertainties.

\begin{figure}
	\includegraphics[width=\columnwidth]{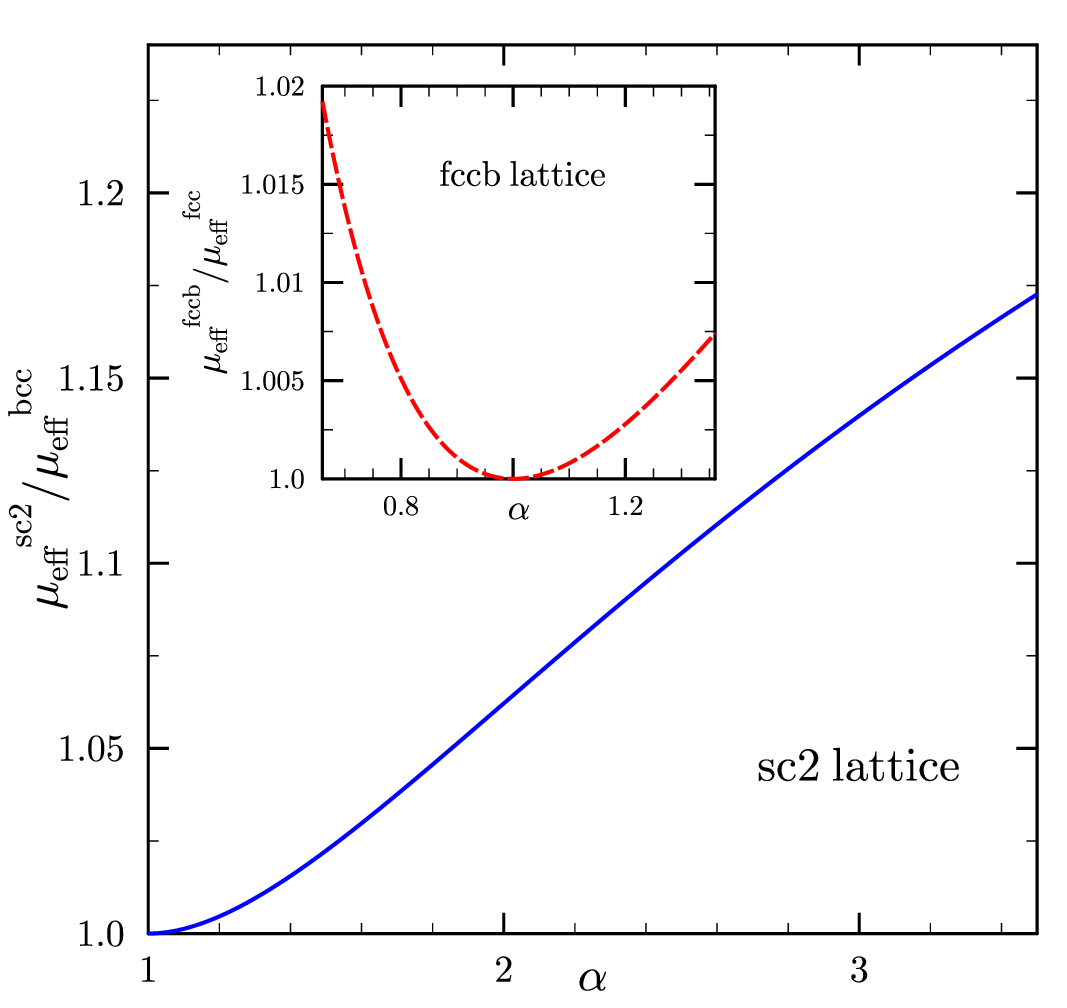}
    \caption{The effective shear modulus of binary Coulomb crystals.}
    \label{fig:mod}
\end{figure}
For astrophysical purposes it is convenient to rewrite $\mu_{\rm eff}^{\rm sc2}$ and $\mu_{\rm eff}^{\rm fc2}$ as a function of concentration of electrons $n_{\rm e}\equiv\bar{Z}n$, where $\bar{Z}$ is the averaged ion charge in the crystal. For the sc2 lattice $\bar{Z}=(Z_1+Z_2)/2$ and for the fc2 lattice $\bar{Z}=(Z_1+3Z_2)/4$ then
\begin{eqnarray}
\mu_{\rm eff}^{\rm sc2}&=&\frac{n_{\rm e}}{a_{\rm e}}
\frac{4(\bar{Z}e)^{2/3}}{(1+\alpha)^2} \\
&\times&  \left(0.0465669(1+\alpha^2)+0.0263234\alpha\right)  \nonumber ~,  \\
\mu_{\rm eff}^{\rm fc2}&=&\frac{n_{\rm e}}{a_{\rm e}} \frac{16(\bar{Z}e)^{2/3}}{(1+3\alpha)^2}  \\
&\times&  \left(0.01848009+0.02276473\alpha+0.07820500\alpha^2\right) \nonumber ~,
\end{eqnarray}
where $a_{\rm e}\equiv(4\pi n_{\rm e}/3)^{-1/3}$. Ratios $\mu_{\rm eff}^{\rm sc2}/\mu_{\rm eff}^{\rm bcc}$ and $\mu_{\rm eff}^{\rm fc2}/\mu_{\rm eff}^{\rm fcc}$ at fixed $n_{\rm e}$ are plotted in Fig. \ref{fig:mod} as a function of $\alpha$, where $\mu_{\rm eff}^{\rm bcc}$ and $\mu_{\rm eff}^{\rm fcc}$ are the effective shear modulus of one-component crystals of ions with charge $\bar{Z}e$.

For the stable lattices changes of ratios in Fig. \ref{fig:mod} are small. For the fully ionized carbon-oxygen mixture in white dwarf envelopes they are equal to 1.01141 for the sc2 lattice and 1.00656 (at $\alpha=4/3$, 25$\%$ C and 75$\%$ O) or 1.00873 (at $\alpha=3/4$, 75$\%$ C and 25$\%$ O) for the fc2 lattice. According to the linear mixing rule the difference between $\mu_{\rm eff}$ of the disordered carbon-oxygen crystal and $\mu_{\rm eff}$ of the one-component crystal with the ion charge $\bar{Z}e$ and at the same $n_{\rm e}$ does not exceed a few percents.

The similar situation takes place for $^{56}$Fe+$^{62}$Ni ($\mu_{\rm eff}^{\rm sc2}/\mu_{\rm eff}^{\rm bcc} \approx 1.00077$ at $\alpha=14/13$) and $^{80}$Ni+$^{120}$Mo ($\mu_{\rm eff}^{\rm sc2}/\mu_{\rm eff}^{\rm bcc} \approx 1.02237$ at $\alpha=1.5$) binary sc2 lattices. Formation of these crystals in the neutron star crust were predicted in \citet{CF16}. They are resistant to the separation into two one-component bcc crystals.

The difference between $\mu_{\rm eff}^{\rm sc2}$ and $\mu_{\rm eff}^{\rm bcc}$ is more important for the oxygen-iron and oxygen-nickel mixtures. For these mixtures $\alpha=3.25$ and $\alpha=3.5$, respectively, and they could form a sc2 lattice. At $\alpha=3.25$ $\mu_{\rm eff}^{\rm sc2}/\mu_{\rm eff}^{\rm bcc} \approx 1.15675$ while at $\alpha=3.5$ $\mu_{\rm eff}^{\rm sc2}/\mu_{\rm eff}^{\rm bcc} \approx 1.17262$. However possibility of formation these crystals in degenerate stars should be checked.

Hence, for the binary crystal mixtures in neutron star crust using the effective shear modulus of a one component Coulomb crystal with averaged charge is seems to be a good assumption.
For compounds with more than two types of ions in the elementary cell it should be checked both analytically (for instance, by the same method which was used in the current paper) and numerically.

~\

~\

{\bf Supplement} 

As I decided not to consider in detail lattices with more than two types of ions in the elementary cell I find it instructive to add a few remarks here.

The validity of the linear mixing rule for the electrostatic energy of the three-component (with three different types of ions) perovskite lattice was checked in \citet{K18}, where it was shown that the difference between the exact result and the result obtained via linear mixing rule is less than $1\%$  \citep[Fig. 1.20 in][]{K18} for all realistic ionic compositions, while the phonon stability of this lattice has never been studied. For other lattices the difference is the same order or less  \citep[see][]{K20}. 

As in \citet{Ch} it was analytically derived that the relation between $\mu_{\rm eff}$ and $U_{\rm M}$ has universal form for any isotropic Coulomb lattice, we can use the linear mixing rule to calculate the effective shear modulus as a Voigt average with the $1\%$ accuracy:
\begin{equation}
\mu_{\rm eff} = -\frac{2}{15} \xi \frac{e^2}{a_e} \sum_{j} Z_j^{5/3}n_j~,
\end{equation}
where sum goes over all types of ions, $\xi$ is the Madelung constant of the one component lattice (the type of the lattice is the same as in mixture).
For the most energetically preferable lattices $\xi \approx -0.896$ \citep[see][]{K18}, which gives
\begin{equation}
\mu_{\rm eff} \approx 0.12 \frac{e^2}{a_e} \sum_{j} Z_j^{5/3}n_j~,
\end{equation}
For studies with precision more than $1\%$ this equation is not suitable and exact equations should be used.

\section*{Acknowledgements}
The author is deeply grateful to D.A. Baiko, A.I. Chugunov and D.G. Yakovlev for help and discussions. This work was supported by Russian Foundation for Basic Research, grant 18-32-20170.

\bsp
\label{lastpage}
\end{document}